\documentclass{iaus}
\usepackage{graphicx}
\usepackage{natbib}

\title[Multiplanet Systems] 
{What to Expect from Transiting Multiplanet Systems}

\author{Daniel C. Fabrycky}
\affiliation{Harvard-Smithsonian Center for Astrophysics\\60 Garden St, MS-51, Cambridge, MA 02138\\ email: {\tt daniel.fabrycky@gmail.com}\\ Michelson Fellow}

\pubyear{2008}
\volume{253}  
\pagerange{???--???}
\jname{Transiting Planets}
\editors{Queloz \& Sasselov, eds.}

\begin{document}

\maketitle

\begin{abstract}
So far radial velocity (RV) measurements have discovered $\sim~25$ stars to host multiple planets.  The statistics imply that many of the known hosts of transiting planets should have additional planets, yet none have been solidly detected.  They will be soon, via complementary search methods of RV, transit-time variations (TTV) of the known planet, and transits of the additional planet.  When they are found, what can transit measurements add to studies of multiplanet dynamical evolution?  First, mutual inclinations become measurable, for comparison to the solar system's disk-like configuration.  Such measurements will give important constraints to planet-planet scattering models, just as the RV measurements of eccentricity have done.  Second, the Rossiter-McLaughlin effect measures stellar obliquity, which can be modified by two-planet dynamics with a tidally evolving inner planet.  Third, TTV is exquisitely sensitive to planets in mean motion resonance.  Two planets differentially migrating in the disk can establish such resonances, and tidal evolution of the planets can break them, so the configuration and frequency of these resonances as a function of planetary parameters will constrain these processes.
\end{abstract}

\firstsection 
\section{Introduction}
This contribution is about systems with at least two planets, at least one of which transits, and it is regretably a ``what to expect'' talk because no such systems are yet known.  That fact, however, ought to be surprising.  In Figure 1 is plotted systems discovered by radial velocity (RV) which display no transits, rank-ordered by the period of the inner planet (just the first 33 shown), and the same plot for the transiting planets made public before the conference.  For systems in which they are known, planetary companions with full RV orbits are plotted on the same line.  The periods covered by the inner planets are comparable between the samples.  But there is a remarkable difference in the number of companions between the samples: 11 companions are known for RV-only planets; none are known for systems with a transiting planet.  Therefore the first expectation we can draw is that the \emph{known} transiting planets have quite a few planetary companions that have until now gone undetected.

\begin{figure}[b]
\begin{center}
 \includegraphics[width=3.0in]{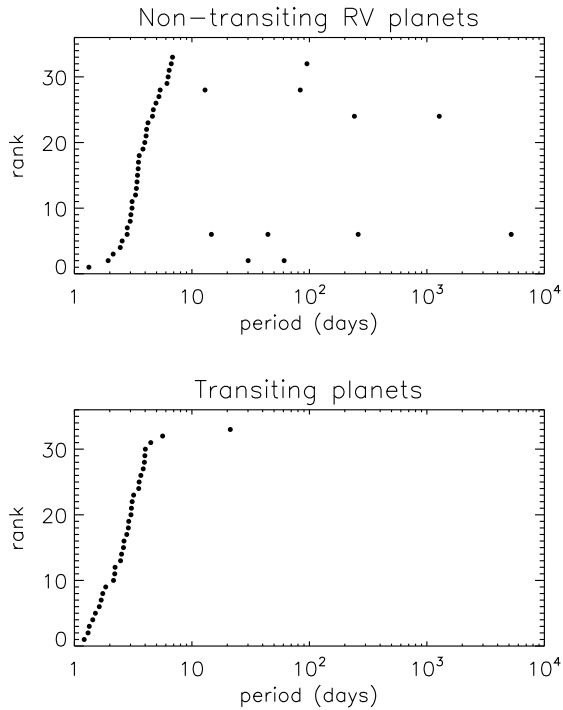} 
 \caption{ The orbital periods of planetary systems (one system per line), rank-ordered by the period of the inner planet, and separated into two groups.  For RV-only systems, only the 33 with the shortest periods are shown, which compare well to the periods of transiting planets, and thus give us baseline expectations for the (yet undetected) companions to transiting planets. }
   \label{fig:pdist}
\end{center}
\end{figure}

	One potential bias for this comparison is that in three of the five systems, the inner planet has an $m \sin i$ more similar to Neptune than to Jupiter, and such planets are not amenable to discovery by transit surveys.  Moreover, those planets may not have even been discovered by RV unless the outer massive planets were intensely observed to constrain their orbits.  Also, lower mass planets may be intrinsically multiple more often (Lovis et al., these proceedings).  On the other hand, dynamical interaction with a transiting planet, as measured by TTV, could in principle reveal companions to much smaller masses than RV.  Instead of trying to unravel such biases, let us proceed with the knowledge that statistics derived from this comparison will only be good at the factor of $\sim 2$ level.

\section{How to Look}
	The simplest way to discover these companions is to monitor the known transiting planets by RV on longer timescales.  A sizable fraction of the transiting planets have published RV observations from only a single few-day observing run; this is not sensitive to companions at all.  Not many additional observations are required, as long as they wisely sample the logarithmic intervals.  Admittedly, the transit-discovered planets are around much fainter hosts than a typical RV planet, so a longer integration time is needed for a decent RV measurement.  However, a multiplanet system with a transiting planet is much more scientifically valuable than, say, several single-planet non-transiting systems which would otherwise be found in the same amount of observing time (\S\ref{sec:dynamics} explores the theoretical value of such multiplanet transiting systems).

\begin{figure}[b]
\begin{center}
 \includegraphics[width=5.0in]{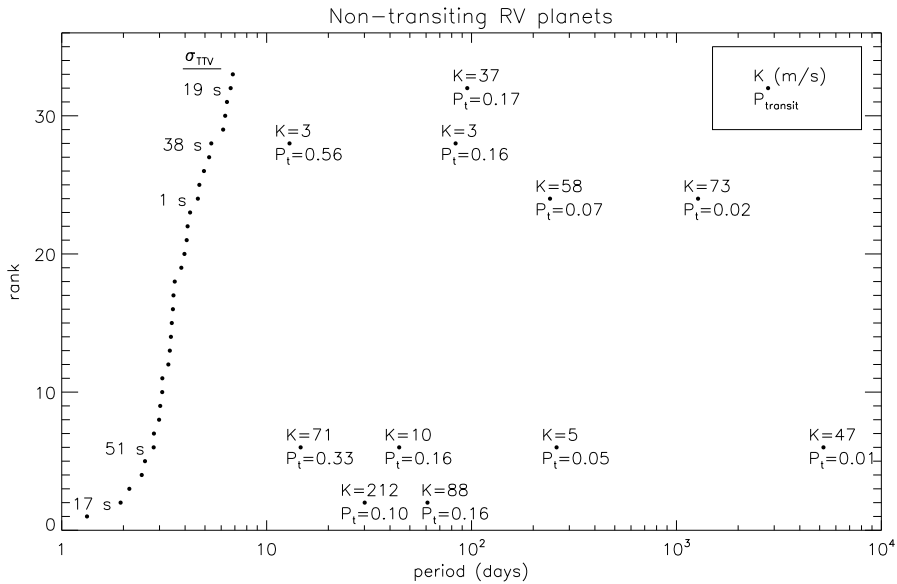} 
 \caption{ The orbital periods of non-transiting RV planetary systems (as in fig. 1), annotated with values to identify what to expect for planetary systems with a transiting planet.  The inner planet of each multiplanet system is labeled with $\sigma_{\rm TTV}$, the amplitude of transit-timing variations induced by the companions, should the planet be found to transit in the future.  The longer-period companions are each given two labels: on top is the RV semi-amplitude (measured), on bottom is the transit probability (under the roughly coplanar hypothesis, assuming the inner planet cuts a random chord across its host).  }
   \label{fig:pdist_annote}
\end{center}
\end{figure}

	In Figure 2 is plotted a rendition of the top panel of Figure 1 in which the RV-only multiplanet systems are annotated in various ways.  If the RV multiplanets are indeed indicative of what to expect for companions of the transiting planets, then this annotation shows how precise the measurements need to be to discover them.  In particular, for RV follow-up, the semi-amplitude $K$ of each of the known companions is given above them.  Seven of the 11 companions have $K>30$~m/s, which is a precision accessible to many groups.
	
	Although RV discovery would be the most straight-forward, there are two promising methods for finding companion planets by photometric means alone.  

	First, the transit-timing method \citep{2005A, 2005HM} is apparently well-publicized, as $\sim 10$ posters at this conference report transit times and refer to how they can rule out Earth mass planets.  This is true if the additional planet is in mean motion resonance, with a large libration amplitude.  For the known RV-only multiplanets of Figures 1 and 2, however, the inner planet (which is the analog of our transiting set) is not in resonance with the other known planets, which limits its TTV signal to only a few tens of seconds, which is difficult to detect with ground-based data.  For these systems, the RMS of deviations from a linear ephemeris (constant period) over five years, $\sigma_{\rm TTV}$, is given in Figure 2 next to the inner planet.  These are ``minimum TTV'' signals assuming the masses of the planets are equal to $m \sin i$ as given in the Extrasolar Planet Encyclopedia, and since these interactions are non-resonant, should be scaled by the unknown $(\sin i)^{-1}$.  Use of the TTV method has resulted in a number of papers quoting impressive upper limits to perturbers in certain nearby or resonant orbits \citep{2005SA, 2007AS, 2008Ma, 2008Mb}.
	
	The second photometric method is looking for the companion planets to transit.  This method has been very under-utilized; the only upper-limit papers are \cite{2007CrollA, 2007CrollB} based on the ``staring'' strategy of MOST (see also the talk on EPOXI by Christiansen et al., these proceedings).  The probable reason for its neglect is that for an individual system a complete search requires good signal-to-noise measurements during the entire period of the putative planet, since the transit signal only lasts for a small fraction of the time $\sim R_\star / (\pi a)$, and such observations are difficult to obtain.  However, we may still wonder why, after all the measurements taken of all the transiting planets, no one has serendipitously found a second planet transiting the star while they were looking.  The probability of it seems not so remote when looking at the gallery of the chords of transiting planets cutting their hosts (Figure 8 of \citealt{2008TWH}), as we are living in the ecliptic plane of those planets.  If companion planets are roughly coplanar with the known ones, we might \emph{expect} to find one soon.  Let us examine this quantitatively.  If we suppose the 5 multiplanet systems in Figure~\ref{fig:pdist_annote} had their inner planets transiting, with uniform distribution in impact parameter, and that the companions are coplanar to the transiting planet to within $i \sim 5^\circ$, then the probability that the outer planet transits is $P_t\simeq a_{\rm in} / a_{\rm out}$.  Each companion in Figure~\ref{fig:pdist_annote} is annotated with this probability.  If we sum this probability over the 8 companions with periods less than 100 days, a difficult but not impossible duration to survey, then $\langle N \rangle = \Sigma P_t = 1.7$.  Here $\langle N \rangle$ is the average number of companion planets we would \emph{expect} to be also transiting the host stars of the known transiting planets.  One method for actually realizing this potential is to combine the datasets of all the transit surveys whose fields cover a given published planet, to probe longer period planets through the increased coverage \citep{2008Fleming}.

\section{ Dynamics of such systems} \label{sec:dynamics}

These multiplanet systems with a transiting planet, which are expected although as-yet undetected, are sure to have dynamical interactions amenable to observation, offering a window into their formation and evolution.  In this section we will examine just a few of the many studies available once such systems are discovered.

\subsection{ Mutual Inclination }

One reason Laplace thought our solar system arose from a spinning disk of matter is that the planets are still in a flattened disk (small mutual inclinations).  However, he also took the small eccentricities of the planets as evidence, and we now know that extrasolar planets show a wide range of eccentricities.  Do the dynamical interactions responsible for the eccentricities also give rise to large mutual inclinations between extrasolar planets?  This question is still open, as the RV signal is insensitive to the inclination or nodal angle of the planet.  Photometric transits famously resolve the inclination problem, giving us line-of-sight inclinations (and thus true masses) for planets.  However, photometric transits are not sensitive to the nodal angle of planets, and thus the photometric signal alone will not reveal whether two planets on Keplerian orbits, transiting the same star, are mutually inclined.  Let us consider systems with two transiting planets to find a way forward.\footnote{Recall that the second planet transiting is one method for finding it in the first place, so assuming that the companion also transits is not too far-fetched.}

A potential solution to this problem is given by spectroscopic measurements during transit: as the occulting disk of the planet traces a path on the rotating host star's photosphere, an anomalous Doppler shift results (the Rossiter-McLaughlin effect; \citealt{2007GW}).  In this way, the angle $\lambda$ between the projected spin axis of the star and the projected orbit normal of the planet is accessible.  Thus, measurements of $\lambda_1$ and $\lambda_2$ of the planets brings the hope of measuring mutual inclination.  Unfortunately a degeneracy $(\lambda, b) \rightarrow (-\lambda, -b)$ spoils the technique for the general case.  However, if $\lambda_1 \approx 0 \approx \lambda_2$, then the two planets must be coplanar to within $\sim 10^\circ$, which should be sufficient to distinguish wildly different planet formation scenarios.

An indirect means of finding the mutual inclination is to measure the impact parameter of the two planets as a function of time.  If the masses of the planets are known, the rate of orbital evolution would give the value of the mutual inclination.  However, one might need a several-year baseline before this secular orbital evolution is measurable \citep{2002M,2008R}.

Finally, a potential direct means for measuring nodal angles (and thus mutual inclination) uses an optical interferometer \citep{2008V}, if one were lucky enough to find a double-transiting system close by.

\subsection{ Tidal Damping and Spin-Orbit Misalignment }

There is a fundamental quantity that is related to planet formation---and was yet another inspiration for Laplace---which can be measured today in one-planet transiting systems: that of stellar spin obliquity (Norio et al., Hebrard et al., these proceedings).  In the last two years theorists have paid attention and started making predictions based upon migration models for hot Jupiters.  In one such class of models, a third body torques or scatters a planet to high eccentricity, which then tidally damps (Wu, these proceedings, \citealt{2007FT}).  Here we wish to point out a secular resonance which can strongly change theoretical predictions if it is not properly modeled.  While the inner planet tidally damps, decreasing in eccentricity and semi-major axis, the frequency at which the inner planet's orbit and stellar spin nodally precess due to their mutual torque continually increases.  Meanwhile, the third body which is responsible for the planet's high eccentricity continues to cause nodal precession on a secular timescale (after strong scattering has finished), but as the inner planet's orbit shrinks, the frequency of this precession continually decreases.  When these two frequencies become comparable, a secular resonance ensues.  The stellar obliquity grows, following an adiabatic invariant.  It is trapped in a stable librational island about the fixed point known as Cassini state 2, the same spin state the Moon inhabits.  Analogous dynamics of the \emph{planet's} spin have been discussed in the literature on transiting planets \citep{2005WH, 2007FJG}.  The final stellar obliquity can move from values near zero to values as high as $\sim90^\circ$ by this mechanism, but probably not to retrograde spins by this mechanism alone.  Rossiter-McLaughlin effect observations of numerous transiting planets are finding a preponderance of well-aligned systems, and more such high-quality measurements will soon put stringent statistical constraints on migration theories.

\subsection{ Tidally Damped Secular Dynamics }

For coplanar two-planet systems, the eccentricities of the two planets torque one another on a secular timescale.  There exists a fixed point to the system in which each planet maintains a constant eccentricity, and the apses precess in synchrony, either aligned or antialigned with each other.  In the case of tidally dissipating inner planets, whose eccentricity would otherwise be expected to damp, \cite{2002WG} have shown analytically that it will first equilibrate at this fixed point.  They also noticed that a single free parameter, $C \equiv k_2 (R_p / a)^5$, controls the value of this eccentricity at a given semi-major axis, where $k_2$ is the tidal Love number which governs the rate of precession induced by tidal distortion of the planet's figure, and $R_p / a$ can be measured precisely in transiting systems (see also Wolf \& Ragozzine, these proceedings).  Therefore, a two-planet system in which the inner planet transits will likely provide our first indirect measurement of tidal distortion of an extrasolar planet.

Such a configuration also allows us to address the concept of pseudo-equilibrium.  Many authors have invoked third bodies to ``pump'' a close-in planet's eccentricity, for planets with a short tidal eccentricity damping timescale, in order to explain a large observed orbital eccentricity or planetary radius.  However, it should be noted that once the tidally dissipating planet reaches the eccentricity fixed point, its semi-major axis shrinks on a longer timescale, as orbital energy is transferred to tidal heat.  This causes the eccentricity of the fixed point to shift, and eventually damp out, on the semi-major axis damping timescale; the fixed point is only a pseudo-equilibrium.  \cite{2007M} has presented a well-developed theory of the evolution of the pseudo-equilbrium, which should be consulted to determine the properties a hypothetical third body would need to solve any given tidal problem.  Figure 3 shows an example of a pseudo-equilbrium: although the system is stable to small perturbations at every configuration, it is secularly evolving.  

\begin{figure}[b]
\begin{center}
 \includegraphics[width=3.0in]{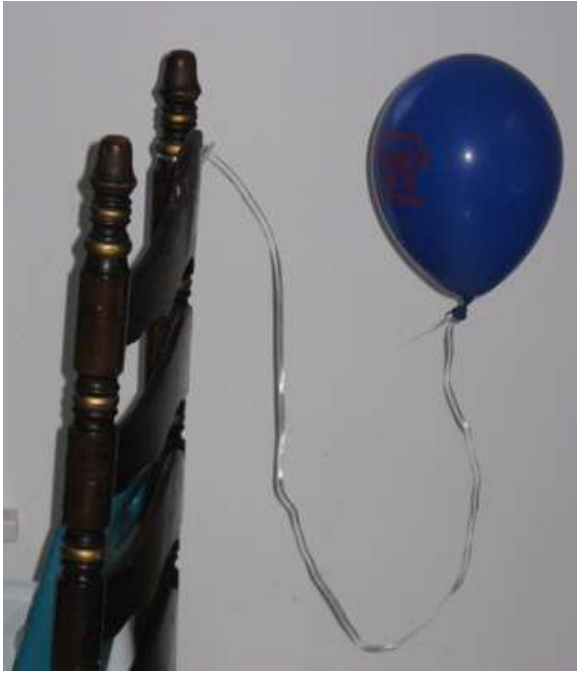} 
 \caption{A helium balloon in pseudo-equilbrium, illustrating the fleeting state of finite eccentricity in a tidally dissipating planet in a two-planet system.  As helium escapes from the balloon, it can no longer hold the burden of the entire ribbon, so it falls to a location in which the chair takes some of the burden.  If the balloon never had the ribbon at all, its transition from the ceiling to the floor would be very fast.  With the ribbon in place, the transition is slower but the outcome is the same.  }
   \label{fig:balloon}
\end{center}
\end{figure}

\subsection{ Resonant Dynamics }

Because of the well-documented sensitivity of TTV measurements to resonant planets, it is clear that such measurements will give excellent constraints on the dynamics of the resonance.  Models of migration of two planets in a gas disk makes certain predictions about which variables librate and at what amplitude (e.g., \citealt{2006BMF}).   Moreover, because of its sensitivity, TTV measurements can provide good completeness when determining the frequency of planets in such resonances, about which planet formation models can also make predictions \citep{2008A}.  When the inner planet is close to the star, its orbit will evolve through tidal dissipation, and most of the known transiting planets are tidally evolved.  Such evolution tends to break mean motion resonances \citep{1980PG, 2007TP}; even their associated fixed points are mere pseudo-equilbria (as in Figure 3).  On the other hand, the anisotropic radiation that close-in planets experience can act as a heat engine on their orbits, moving inner planets to larger semi-major axis, which can sweep exterior planets into resonance \citep{2008F}.

\section{ Conclusions }

The main expectations we have for multiplanet systems with at least one transiting planet are (1) that they exist and should be plentiful among the current crop of transiting exoplanets, (2) there are already methods in use---RV, TTV, transits---that will find them, probably soon, and (3) such systems will be enormously powerful probes of planetary formation and evolution: the theorists are salivating.

\section{Acknowledgments}

These ideas were developed during a Michelson Fellowship supported by the National Aeronautics and Space Administration and administered by the Michelson Science Center.  I thank Avi Loeb for pointing out that transits of companions (or a lack thereof, once it becomes statistically implausible) can constrain the typical mutual inclination.

\bibliography{fabrycky} \bibliographystyle{apj}

\end{document}